\documentclass[a4paper,UKenglish,cleveref, autoref, thm-restate]{lipics-v2021}

\pdfoutput=1
\hideLIPIcs

\graphicspath{{./figs/}}

\title{Automated Rendering of Multi-Stranded DNA Complexes with Pseudoknots}

\author{Ma{\l}gorzata Nowicka}{Department of Computer Science and Department of Media, Aalto University, Finland}{malgorzata.nowicka@aalto.fi}{https://orcid.org/0009-0009-2531-9820}{}

\author{Vinay K. Gautam}{Department of Computer Science, Aalto University, Finland \and Department of Chemical Engineering, NTNU, Norway}{vinay.k.gautam@ntnu.no}{https://orcid.org/0000-0003-1506-2071}{}

\author{Pekka Orponen}{Department of Computer Science, Aalto University, Finland}{pekka.orponen@aalto.fi}{https://orcid.org/0000-0002-0417-2104}{}

\authorrunning{M. Nowicka and V.\,K. Gautam and Pekka Orponen}

\Copyright{Ma{\l}gorzata Nowicka, Vinay K. Gautam and Pekka Orponen}


\category{}

\funding{Research supported by Academy of Finland grant 311639, ``Algorithmic Designs for Biomolecular
Nanostructures (ALBION)''.}

\nolinenumbers

\newcommand{\sign}{\text{sign}}

\begin{document}

\maketitle

\begin{abstract}
We present a general method for rendering representations of multi-stranded DNA complexes from textual descriptions into 2D diagrams. The complexes can be arbitrarily pseudoknotted, and if a planar rendering is possible, the method will determine one in time which is essentially linear in the size of the textual description. (That is, except for a final stochastic fine-tuning step.) If a planar rendering is not possible, the method will compute a visually pleasing approximate rendering in quadratic time. Examples of diagrams produced by the method are presented in the paper.
\end{abstract}

\section{Background}
\label{sec:background}

Multi-stranded DNA complexes are of wide interest in the area of DNA nanotechnology. In the present work, we focus specifically on their use as building blocks in DNA strand displacement (DSD) systems~\cite{zhang}, which are a widely-used methodology for embedding computational processes in the interactions of pre-designed short DNA strands. 

The design and modelling of DSD systems is typically done at the binding-domain level, where a \emph{domain} is a contiguous and functionally distinct sequence of nucleotides. A DNA \emph{strand} is then a sequence of connected domains. Strands are oriented, starting at the \emph{5'-end} and ending at the \emph{3'-end}. Domains of different strands may bind with each other in antiparallel orientation, forming a DNA complex which is (mathematically speaking) a multiset of strands connected by domain bindings. At the design phase one is however generally not interested in individual strands and complexes, but rather distinct similarity types (equivalence classes) of these. A similarity type of DNA complexes is called a DNA \emph{species}, whereas with some abuse of terminology the term ``strand'' is retained also for similarity types of strands.

\begin{figure}
\centering
  \includegraphics[width=0.9\linewidth]{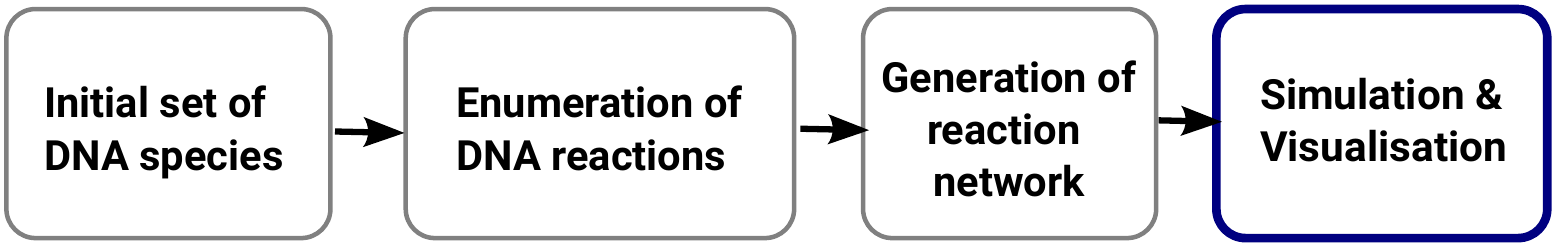}
  \caption{DSD system design pipeline.}
  \label{pipeline}
\end{figure}

Automated tools for designing and modelling DSD systems commonly follow a four-stage pipeline (Figure~\ref{pipeline}) in which we are presently concerned with the final step of visualisation --- or more specifically the task of producing 2D diagrams of species generated from an initial set of species. Such diagrams are highly useful to understand the characteristics of species emerging in a given DSD system, validate the system design, and communicate related results to a broader audience.

\begin{figure}
\centering
  \includegraphics[width=0.6\linewidth]{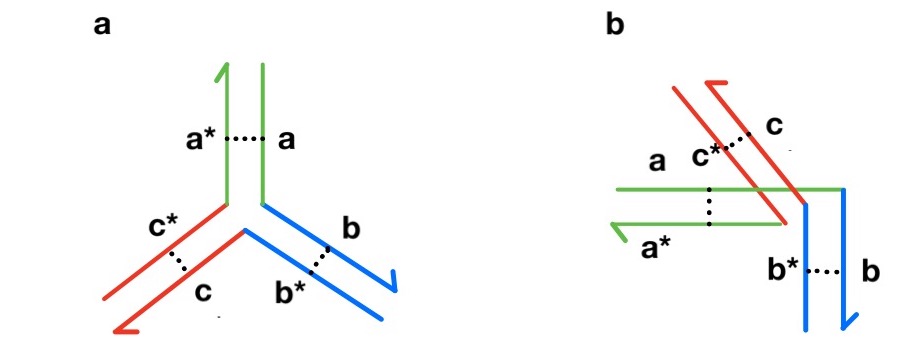}
  \caption{DNA species diagrams. a: Good drawing. b: Unsatisfactory drawing.}
  \label{drawing-example}
\end{figure}

As is customary in the DSD design literature~\cite{zhang}, we aim to render multi-strand DNA species as 2D diagrams, where each strand is represented as a sequence of contiguous line or curve segments that depict the domains, and domain pairings are indicated by some connector symbol, for instance a line segment connecting the domain midpoints~(Figure~\ref{drawing-example}). Ideally, a 2D rendering of a given DNA species should satisfy the following conditions: (i)~the domain depictions are straight-line segments of equal length; (ii)~bound domains are drawn close to each other, ideally at a fixed distance and in antiparallel orientation; (iii)~the polylines (or polycurves) representing species do not cross each other in the drawing plane. For structurally complex species, the task of satisfying all these constraints simultaneously and satisfactorily can be quite difficult or even impossible.

\paragraph*{The challenge of pseudoknots}

\begin{figure}
\centering
  \includegraphics[width=0.7\linewidth]{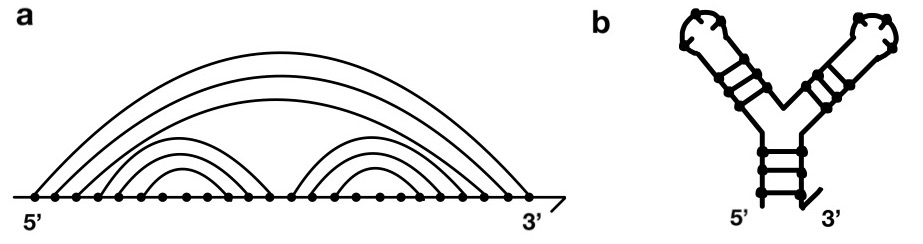}
  \caption{RNA secondary structure diagrams. a: A 1D arc diagram depicting base pairings. b:~Corresponding 2D secondary structure diagram.}
  \label{rna-structure}
\end{figure}

Many current DNA species rendering algorithms take inspiration from methods used in visualising RNA secondary structures. In RNA, a single strand folds upon itself and forms bindings pairing its nucleotides. In simple ``nested'' pairing arrangements, this process results in a tree-like secondary structure with contiguous \emph{stem} segments of paired bases constituting the edges and \emph{loops} of unpaired bases forming the leaves (Figure~\ref{rna-structure}). Visualisation methods for such structures commonly follow this intrinsic tree layout~\cite{bruccoleri, shapiro}.

This approach extends to DNA visualisations as well, because if the strands constituting a multi-strand DNA species are arranged in some linear order and connected into a single strand, the rendering problem reduces to the corresponding one for RNA secondary structures. Thus, the only additional task is to determine a good ordering for the strands. (Of course for $n$ strands there are $n!$ possible orderings, so the task is not quite straightforward.) This method is widely used in visualisations of nested DNA species, for instance in the DyNAMiC Workbench tool~\cite{grun}.

\begin{figure}
\centering
  \includegraphics[width=0.7\linewidth]{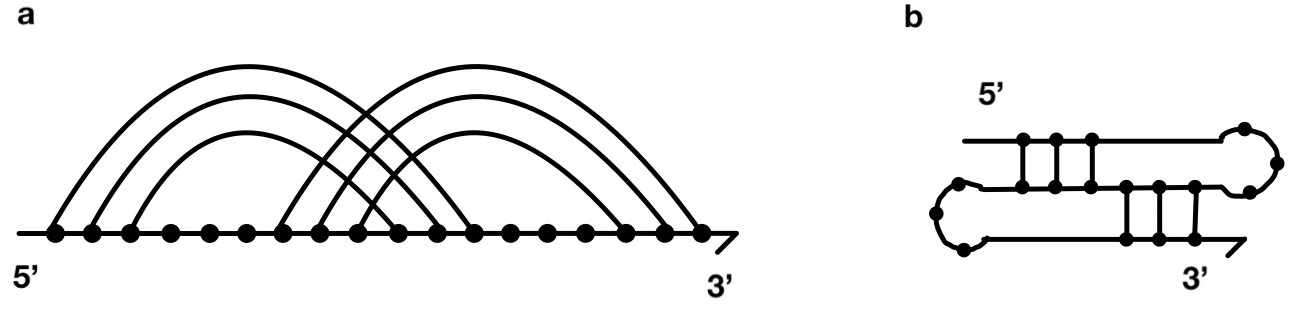}
  \caption{An H-type RNA pseudoknot. a: Arc diagram. b: Secondary structure diagram.}
  \label{h-pseudoknot}
\end{figure}

A problem emerges, however, when base pairings are not nested, but cross each other in the linear strand arrangement, as in Figure~\ref{h-pseudoknot}a. In this \emph{pseudoknotted} situation, the secondary structure no longer has a tree-like character (Figure~\ref{h-pseudoknot}b), and many challenges ensue both in diagramming and otherwise analysing the structure~\cite{akutsu, lyngso}.

A number of RNA visualisers nevertheless support the rendering of pseudoknotted structures using alternative approaches. PseudoViewer3~\cite{byun} identifies and draws known types of pseudoknots, and arranges the rest of the structure in relation to them. The tool jViz.Rna 2.0~\cite{wiese} uses a force-directed algorithm for assembling the nucleotides, whereas jViz.RNA~4.0~\cite{shabash} and forna~\cite{kerpedjiev} identify the pseudoknotting connections and draw them on top of a pruned pseudoknot-free skeleton of the species. 

A multi-strand DNA species can be considered pseudoknotted, when no linear ordering of its constituent strands results in a nested base-pairing arrangement. Rendering of pseudoknotted species in DSD systems has been little studied, because when there is no tree structure to be identified, exploring the space of $n!$ permutations for a species of $n$ strands becomes challenging. VisualDSD~\cite{lakin}, a widely-used tool for designing and modelling DSD systems, takes the approach of rendering a pseudoknot-free skeleton of the species and then just highlighting  the pseudoknot-inducing domains, without connecting them visually. While in RNA this approach may be helpful in order to easily identify the type of the pseudoknot, in DNA it is less useful in illustrating the structure. (For examples, see Figure~\ref{examples}.) A force-directed approach was used in~\cite{gautam} but it covers only relatively simple pseudoknotted structures. Other common DSD system design tools such as NUPACK~\cite{zadeh} and DyNAMiC Workbench~\cite{grun} do not provide options for visualising pseudoknots.

\section{A DNA species rendering algorithm}

We approach the task of rendering multi-strand DNA species in a Gordian-knot manner: since a key target is to achieve a noncrossing arrangement for the strands, we aim to achieve this directly by a planar graph embedding technique that is guaranteed to find such an arrangement efficiently if one exists. For cases where no planar (noncrossing) representation is possible, we present an efficient and visually appealing approximate method.

Luckily, quite many pseudoknotted (RNA and multi-strand DNA) structures have planar renderings: a simple example is the H-type pseudoknot in Figure~\ref{h-pseudoknot}, for which a planar drawing is presented in subfigure~\ref{h-pseudoknot}b. In RNA, planar pseudoknots are closely related to the Rivas-Eddy class of secondary structures, which is also one of the largest known algorithmically tractable families of RNA structures~\cite{gao,lyngso-pedersen,rivas}.

\begin{figure}%[h!]
\includegraphics[width=0.95\linewidth]{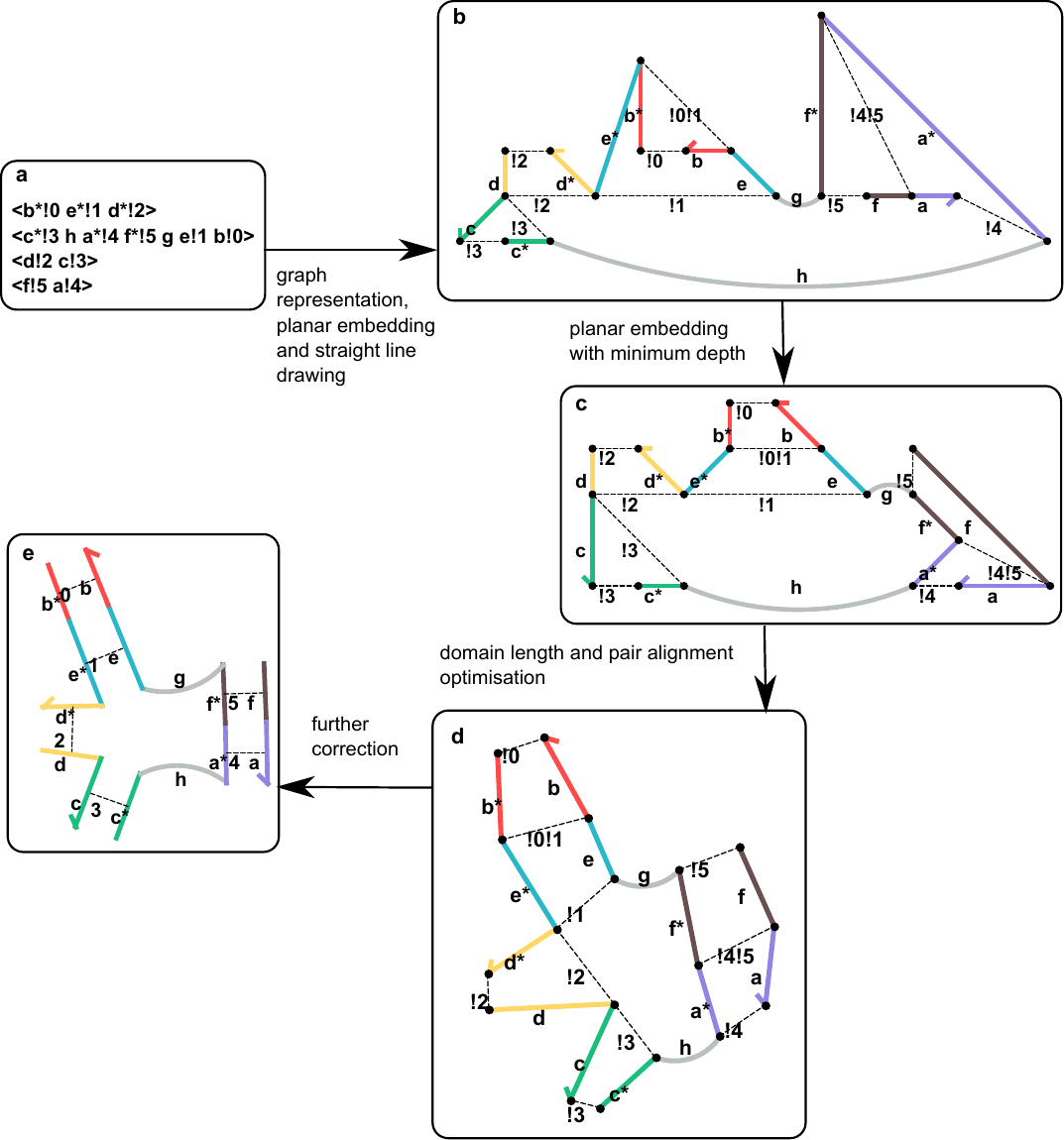}
  \caption{Rendering steps of a species $S$.
      Each box displays a stage of the rendering algorithm. The captions on the arrows indicate the operations taken to achieve the next step. a: Text representation. b: Straight-line drawing. Vertices mark the ends of domains, solid lines represent the domains and dashed lines the pair connections. c: Minimised depth. d: Initial optimisation. e: Final drawing.}
      \label{rendering}
      \end{figure}
      
\paragraph*{Text representation}
For textual description of multi-strand DNA species we use a 'process calculus' notation based on named-domain pairings and introduced to the DSD context in~\cite{petersen} (see Figure~\ref{rendering}a). Lowercase letters and numbers are used as domain names (e.g.\ $a$).
A complementary domain has the same name with an additional asterisk appended (e.g.\ $a^*$). A domain with a given name can appear repeatedly. A pairing of domains is indicated by an exclamation mark followed by a matching index, appended to the domain names (e.g.\ $a!1$, $a*!1$). Each pairing must be unique.
A strand comprises one or more sequentially connected domains, enclosed in angle brackets: $\langle a \ b \rangle$. A species is a collection of strands, separated by two forward slashes: $\langle a \ b\rangle //\langle c\rangle $. Paired domains must belong to the same species in the description.

\paragraph*{Graph representation and algorithm generality}

Figure \ref{rendering}a presents an example of a textual description of a species $S$ and Figure \ref{rendering}b its  graph representation $G$. The vertices in $G$ correspond to the domain boundaries, including the 5'- and 3'-ends, of the strands in $S$. The domains within the strands (solid lines in Figure \ref{rendering}b) and the domain pairings connecting the strands (dashed lines in Figure \ref{rendering}b) constitute the edges in $G$. To ensure the antiparallel orientation of each domain pairing $\{a, a^*\}$, the 5'-end of $a$ is connected to the 3'-end of $a^*$ and vice versa. We say that a species $S$ is \emph{planar} if $G$ is a planar graph, i.e.\ can be embedded on a plane without edge crossings. Determining if $S$ is planar can be done in linear time by using any planarity testing algorithm, e.g.\ the Boyer-Myrvold edge addition method~\cite{boyer-myrvold}. 

To characterise the generality of the algorithm we consider the notion of \emph{book thickness} of a graph~\cite{bernhart} that is widely used in the context of RNA theory~\cite{haslinger}. For a given graph $G$ a \emph{book embedding} comprises a linear ordering of the vertices of $G$, called the book \emph{spine}, and a partition of its edges. The edges in each class of the partition are drawn as non-crossing arcs on a separate halfplane (a \emph{page}) bordered on the book spine. The book thickness of $G$ is then the minimum number of pages needed to achieve such a book embedding, for some ordering of the vertices along the spine. It is known that the book thickness of a graph $G$ is at most 2 if and only if $G$ is a subgraph of a planar Hamiltonian graph~\cite{bernhart}.

In RNA the vertices represent nucleotides and the edges their pairings~\cite{haslinger}. There is only one possible ordering of the vertices, and in case of linear RNA, connecting the first and the last vertex makes the graph Hamiltonian. Therefore the existence of a planar rendering of an RNA secondary structure is equivalent to the condition that its arc diagram can be embedded on at most 2 pages. This can be tested by e.g.\ checking the bipartiteness of an inconsistency graph $\theta(V, E)$~\cite{haslinger}, where every vertex corresponds to a crossing of the arcs in the arc diagram. 

In multi-strand DNA, the ordering of the vertices is not given, which makes the problem of determining if a 2-page embedding exists NP-complete~\cite{chung, wigderson}. Moreover, there exist maximal planar graphs that are not Hamiltonian, therefore some planar graphs might need more than 2 pages (see Figure~\ref{examples}c). In fact, any planar graph can be embedded on 4 pages~\cite{kaufmann, yannakakis}. Therefore the notion of planar DNA species as defined in this paper encompasses all 1-page (= nested) species, all 2-page pseudoknots, and some pseudoknots that require more than 2 pages and are planar.

\paragraph*{Straight-line drawing}

It is known that any planar simple graph (= no multi-edges, no self-loops) has a plane embedding with straight-line edges~\cite{wagner, fary, stein}. By definition, a species graph $G$ as described earlier is always simple, and can thus be drawn with straight lines if it represents a planar species. Rendering $G$ this way satisfies the conditions that the domains are drawn as straight-line segments and the shortest distance between domains within a pair is represented by a straight line. 

Given a planar species graph $G$, a drawing (plane embedding) $D$ of $G$ can be achieved in linear time by an algorithm described in~\cite{chrobak} that produces crossing-free, straight-line drawings on an $(n-2)\times(n-2)$ grid for graphs with $n$ vertices. The algorithm consists of four stages. 
\begin{enumerate}
    \item $G$ is augmented by additional edges to achieve the  biconnectivity required for the canonical decomposition \cite{harel} described in step 3. 
    \item A combinatorial embedding of $G$ (clockwise ordering of incident edges at each vertex) is determined e.g.\ with the Boyer-Myrvold method~\cite{boyer-myrvold}. The face of $G$ with most vertices is chosen as the external face.
    \item The canonical decomposition is computed. This is an ordered partition of the set of vertices of $G$ and determines the order of the vertices in the process of drawing.
    \item The graph is drawn on a plane according to the algorithm in \cite{chrobak}. The augmented edges are removed from $D$. 
\end{enumerate}

All stages of the algorithm are of linear time complexity. An example result is presented in Figure~\ref{rendering}b.

If $G$ is not planar, a technique for generating orthogonal drawings (in orthogonal drawings the edges may bend 90 degrees) for any graph called topology-shape-metrics \cite{battista, tamassia} is used. It consists of three stages.

\begin{enumerate}
    \item $G$ is planarised and a planar embeddding is computed. The planarisation process~\cite{gutwenger} first computes planar subgraphs of $G$ and then reinserts the remaining edges, trying to minimise the number of crossings and creating a dummy vertex in the place of each crossing. The embedding is determined in the same way as in step 2 of the previous method.
    \item In the orthogonalisation stage the bends of the edges are determined with a bend minimisation approach described in~\cite{tamassia-bend}.
    \item In the compaction stage the lengths of the edges are computed and the coordinates of the vertices are determined. In the final rendering of $D$, the edge bending is ignored and edges are represented with straight lines. This may lead to crossings in $D$. 
\end{enumerate}

The complexity ranges from $O(n)$ to $O(k^2(n+c)^2\log(kn))$ depending on the planarity of the graph, number of crossings, orthogonalisation and compaction method ($c$ is the number of crossings and $k$ is a parameter related to the tidy expansion step, all described in~\cite{tamassia}).
 
\paragraph*{Minimum depth and maximum external face drawing}

The \emph{depth} of a graph is a measure of the topological nesting of its biconnected components (maximal biconnected subgraphs)~\cite{pizzonia}. In a species graph $G$, the biconnected components comprise connected domains and domain pairs. When depth is not minimised, some domains are topologically nested inside other pairs, which prevents the outer pair from achieving the desired proximity between its domains. 

\begin{figure}
\includegraphics[width=0.9\linewidth]{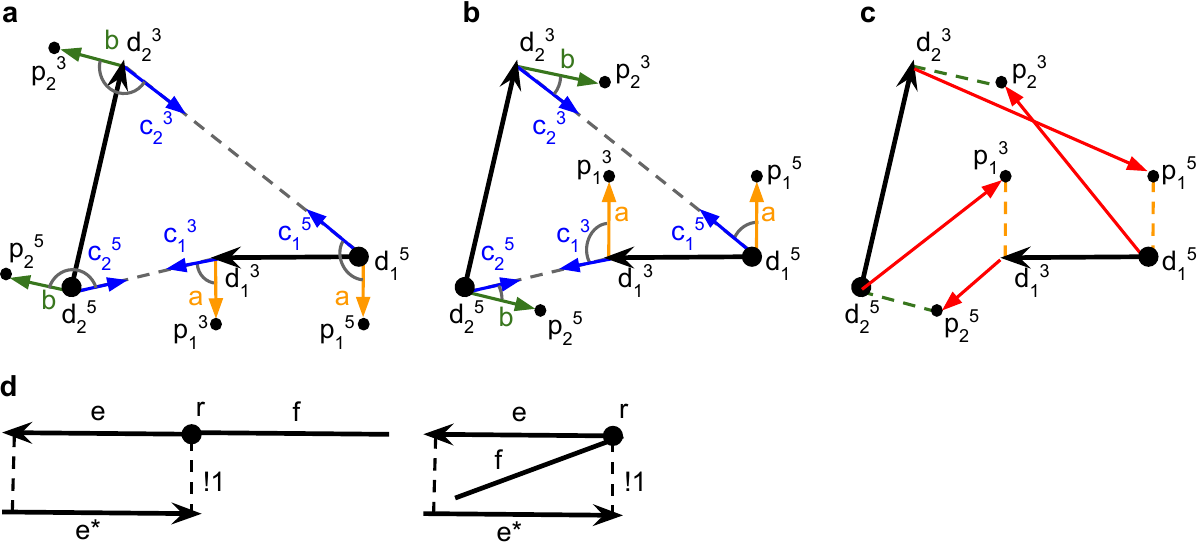}
  \caption{Computing the correct placement of $d_1$ and $d_2$ and choosing the correct embedding for the vertex $r$.}
      \label{figure-sa}
      \end{figure}

Consider a planar embedding of a graph of a species with strands \linebreak $\langle e!1 \ f\rangle  \ \langle e*!1\rangle $. Let $r$ be a vertex between domains $e$ and $f$ and $!1$ be the edge connecting the bounded domains. There exist two possible combinatorial embeddings in $r$ (see Figure~\ref{figure-sa}d) but only the one of minimum depth will result in a drawing that minimises the distance between the paired domains. In Figure \ref{rendering}b, pairs $aa^*$ $bb^*$, and $ff^*$ are topologically nested, and therefore pairs $cc^*$, $dd^*$, and $ee^*$ cannot properly align.

In order to minimise the depth, a linear time embedding algorithm~\cite{gutwenger-depth} that uses a SPQR-tree data structure is applied to the second stage of the planar $G$ drawing. It chooses a planar embedding of maximum external face size among all planar embeddings with minimum depth. Embeddings with maximum external face have shown to improve certain metrics like the length of the edges, size of the area \cite{gutwenger2}. The result of the new embedding is presented in Figure \ref{rendering}c.

\paragraph*{Drawing optimisation}

The final step for both planar and non-planar $G$ is to adjust the lengths of the edges without introducing new crossings in $D$.  This consideration imposes an additional requirement on force-directed graph layout methods commonly used for such purposes. In a force-directed layout algorithm, the connected vertices are attracting each other until the distance between them is of desirable length, and the unconnected vertices are repelling each other. An additional node-edge repulsion formula tries to prevent the overlapping of nodes and edges, but due to the discreteness of the movements of the vertices, new crossings may appear. Article~\cite{bertault} introduces the use of zones around every vertex that confine its movement at each iteration of the algorithm and prevent the formation of new crossings (see Figure \ref{rendering}d).

Further correction in the placement of the domains is required, as the previous step does not achieve the desired lengths for all the edges or the right alignment between domains in a pair. A simulated annealing algorithm~\cite{bertsimas} is used. Let domains $d_1$ and $d_2$ be paired and represented by vectors, such that $d_1=[d_1^{3}, d_1^{5}]$ and $d_2=[d_2^{3}, d_2^{5}]$. A distance to the correct placement of  $d_1$ and $d_2$ is computed in three steps as follows (see Figure~\ref{figure-sa}).

\begin{enumerate}
    \item Let $a$ and $b$ be vectors rotated 90 degrees counter-clockwise in relation to $d_1$ and $d_2$ respectively. Let $p_1^3$ and $p_1^5$ be the endpoints of $a$ anchored at both ends of $d_1$ and $p_2^3$ and $p_2^5$ be the endpoints of $b$ anchored at both ends of $d_2$. Let $c_1^3 = [d_1^3, d_2^5]$, $c_1^5 = [d_1^5, d_2^3]$, $c_2^3 = [d_2^3, d_1^5]$, and $c_2^5 = [d_2^5, d_1^3]$ (Figure~\ref{figure-sa}a). 
     \item The correct side for $d_1$ to bind with $d_2$ is determined by the inequality: 
    \[\sign(\cos(a, c_1^3))+\sign(\cos(a, c_1^5))+\sign(\cos(b, c_2^3))+\sign(\cos(b, c_2^5)) \geq 3 \]
    where $\sign(x) = 0$ if $x \leq 0$, and $1$ otherwise.
    If the inequality is not satisfied, rotate $a$ and $b$ so that they are in 90 degrees clockwise relation to $d_1$ and $d_2$ respectively, as shown in Figure \ref{figure-sa}b. Note the change of the position of the endpoints of $a$ and $b$. 

    \item Compute the distances from $d_2^5$ to $p_1^3$, $d_2^3$ to $p_1^5$, $d_1^5$ to $p_2^3$, and $d_1^3$ to $p_2^5$ as in Figure \ref{figure-sa}c.
\end{enumerate}

At each step of the algorithm, three factors are being optimised: the distance to the correct placement of domains for all pairs, the deviation from the desired lengths of the edges, the deviation from the desired distance between the ends of loop domains.

Figure \ref{rendering}e presents the final outcome where, to follow standard practice, the double vertices at the domain boundaries have been removed and replaced by single edges connecting the respective paired domains. 

\paragraph*{Implementation}

The planar embeddings of minimum depth and maximum external face for both planar and non-planar species and the graph drawing were achieved with the help of algorithms in the Open Graph Drawing Framework \cite{ogdf}. The visualisation is implemented as a part of the XDSD design tool~\cite{xdsd}.

\section{Results}

\begin{figure}
\centering
\includegraphics[width=0.9\linewidth]{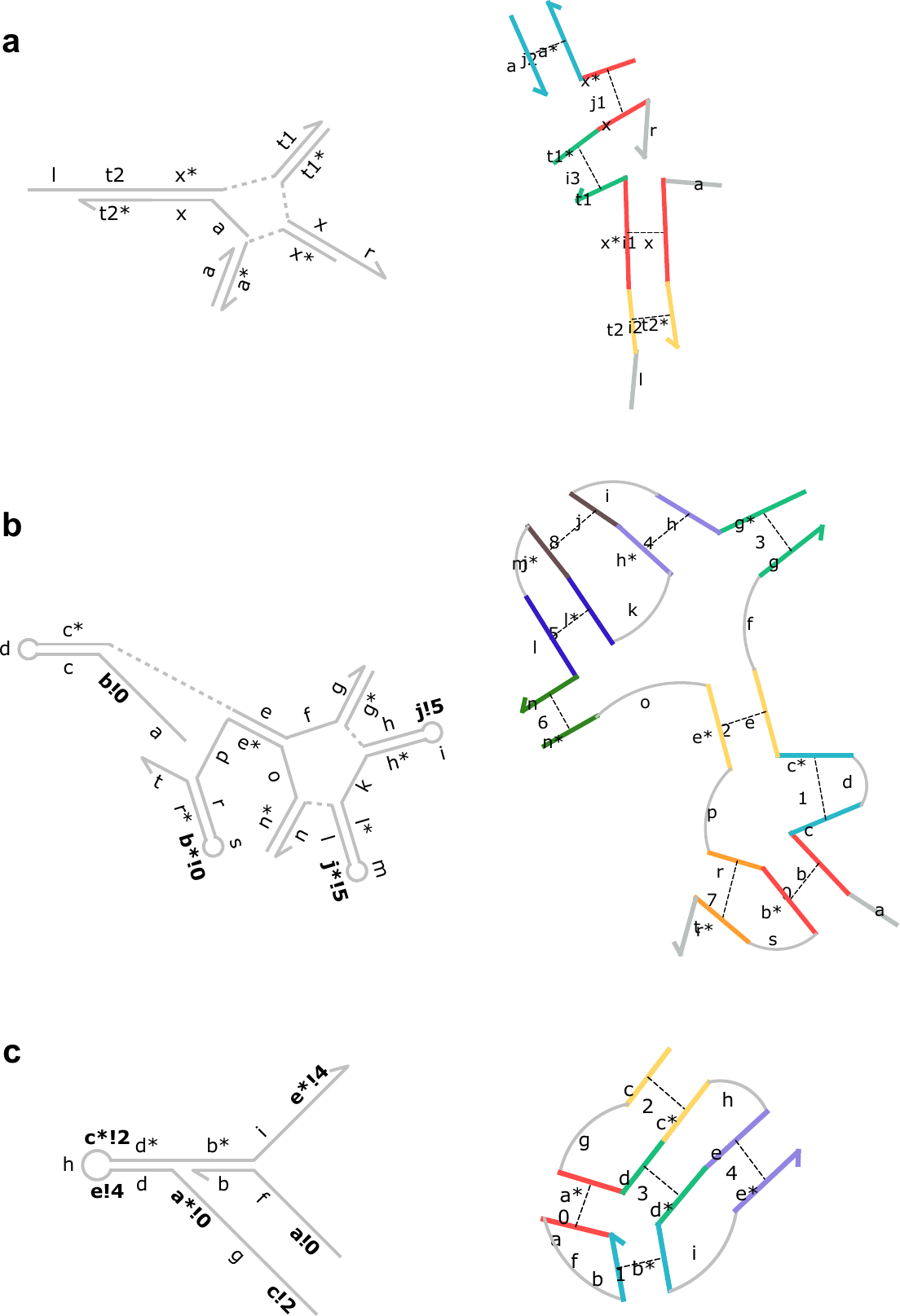}
  \caption{Three species renderings generated by VisualDSD on the left and our algorithm on the right.
      }
      \label{examples}
\end{figure}

Figure \ref{examples} demonstrates the visualisation possibilities of the algorithm. The results are compared with VisualDSD~\cite{lakin}, a widely-used tool that accepts pseudoknotted multi-strand designs as inputs.

The rendering algorithm in VisualDSD admits multi-strand, pseudoknotted species with loops as inputs. The bonds that make the species a pseudoknot (pseudoknotting connections) are accepted but not specifically handled except for denoting them by boldening the domain and bond name.

Figure~\ref{examples}a presents a 1-page, nested species. The VisualDSD method does not maintain the connectedness of the strands and instead draws dashed lines between the separated domains. Our algorithm resolves to manipulating the lengths of the domains in order to keep the domains together.

The species in Figure~\ref{examples}b is a 2-page pseudoknot. The VisualDSD version  breaks the species apart and superposes the pseudoknotting connections after the other parts of the species is rendered. Our method does not distinguish between the regular and pseudoknotting connections and renders all of them.

Figure~\ref{examples}c shows another pseudoknot, a 3-page species that has a planar representation. Again, the VisualDSD rendering ignores the pseudoknotting connections. These noticeably change the appearance of the species, once the pseudoknotting domains are brought together.

\section{Conclusions}

\paragraph*{Discussion}
Based on notions of planar pseudoknots and planar multi-strand DNA species,  we have presented a rendering method that is able to closely follow the conventions of DSD system drawings (strands do not cross each other, domains are represented as fixed length straight-line segments, and bounded domains are close to each other at a fixed distance) while keeping the strands connected and pseudoknots visualised.

In renderings for pseudoknot-free species, conventionally there is one binding side for all the domains in the species (e.g. clockwise or to the right w.r.t.\ the 5'-3' direction of the domains). Our algorithm on the other hand does not distinguish between pseudoknotted and pseudoknot-free species, and automates the embedding of the species without paying attention to the consistency of the binding sides (see Figure \ref{figure-sa}a). This may in some cases lead to decreased readability of the drawing.

\paragraph*{Further work}

The current implementation of the embedder cannot be applied before the augmentation stage of the drawing algorithm. Therefore the augmented edges are embedded as well, which after their removal may result in a species rendering with crossings and non-minimal depth. In a future version, one could aim to apply the embedder before the graph is augmented, so that the addition and removal of the support edges does not affect the depth of the resulting graph.

The drawing optimisation now uses a simulated annealing approach with fixed hyperparameters that does not take into account the production of new crossings. This way the optimisation process is faster, but may exert too much or too little pressure on the species, resulting result in new crossings or unaligned pairs of uneven-length domains respectively. Further developments of the drawing optimisation part are needed since this is the only non-linear step in the algorithm.

\bibliography{dsdrender}

\end{document}